\documentclass[aps,pre,11pt,nofootinbib,byrevtex,showkeys,longbibliography,notitlepage,preprintnumbers]{revtex4-1}
\usepackage{microtype}
\usepackage{amssymb}
\usepackage{amsmath}
\usepackage{amsthm}
\usepackage{enumerate}
\usepackage{booktabs}
\usepackage{color}
\definecolor{myblue}{rgb}{0.153,0.322,0.706}
\usepackage[colorlinks,linkcolor=myblue,urlcolor=myblue,citecolor=myblue,breaklinks=true]{hyperref}

\newcommand{\be}{\begin{equation}}
\newcommand{\ee}{\end{equation}}
\newcommand{\reals}{\mathbb{R}}
\newcommand{\ra}{\rightarrow}
\newcommand{\p}{\partial}
\newcommand{\cL}{\mathcal{L}}
\newcommand{\cH}{\mathcal{H}}
\newcommand{\id}{1\!\! 1}
\newcommand{\eps}{\varepsilon}
\newcommand{\exdiff}[1]{\textrm{[#1]}}
\newcommand{\myemph}{\textbf}
\renewcommand{\emph}{\textit}

\begin{document}

\title{Introduction to dynamical large deviations of Markov processes$^*$\footnote[0]{$\!\!\!^*$Lecture notes for the 2017 Summer School on Fundamental Problems in Statistical Physics XIV, 16-29 July 2017, Bruneck (Brunico), Italy.}}
\author{Hugo Touchette}
\affiliation{National Institute for Theoretical Physics (NITheP), Stellenbosch 7600, South Africa}
\affiliation{Institute of Theoretical Physics, Department of Physics, Stellenbosch University, Stellenbosch 7600, South Africa}

\date{\today}

\begin{abstract}
These notes give a summary of techniques used in large deviation theory to study the fluctuations of time-additive quantities, called dynamical observables, defined in the context of Langevin-type equations, which model equilibrium and nonequilibrium processes driven by external forces and noise sources. These fluctuations are described by large deviation functions, obtained by solving a dominant eigenvalue problem similar to the problem of finding the ground state energy of quantum systems. This analogy is used to explain the differences that exist between the fluctuations of equilibrium and nonequilibrium processes. An example involving the Ornstein-Uhlenbeck process is worked out in detail to illustrate these methods. Exercises, at the end of the notes, also complement the theory.
\end{abstract}

\maketitle

\section{Introduction}

My aim in these notes is to give an introduction to large deviation techniques used to calculate the probability distribution of physical quantities, called \textit{dynamical observables}, which are time-integrated functionals of Markov processes. These observables have come to play in the last years an important role in the context of nonequilibrium systems, and the recent discovery of fluctuation symmetries (also called fluctuation relations) generally satisfied by these systems. They also appear naturally when defining energy-like quantities, such as work and heat, in the context of noisy systems modelled by Markov processes, a field of research commonly referred to now as \textit{stochastic thermodynamics} (see Seifert's contribution to this volume of lecture notes).

The results covered are known in large deviation theory, but are not widespread, certainly not in statistical physics, and cannot be found all put together in a single reference. The invitation to lecture at the FPSP School (to which I participated as a student in 2005) offers a good opportunity to summarize them by appealing to what physics students know best: quantum mechanics.

The link with quantum mechanics is a natural one because all large deviation calculations related to time-additive functionals of Markov processes reduce, when one does not consider the low-noise limit, to an eigenvalue calculation, which is very close in spirit to finding the energy levels of a quantum system and, particularly, its ground state energy. For this reason, it is not surprising to see many techniques of quantum mechanics (such as the Bethe ansatz and the density matrix renormalization group) being applied to study the fluctuations of nonequilibrium processes. In a way, if you know quantum mechanics, you also know Markov processes. 

Compared to quantum systems, however, there is a fundamental difference that arises when dealing with Markov processes, namely, that the operator or matrix that we must consider to study their steady state and fluctuations is, in general, not Hermitian. This raises many technical but important questions, which are precisely the questions that are hard to find in the literature, such as: 
\begin{itemize}
\setlength\itemsep{0.1em}
\item What can we say in general about the spectrum of non-Hermitian operators? 
\item Under what conditions is that spectrum real?
\item Which function space must we use to solve the eigenvalue problem? 
\item What are the boundary conditions for the eigenfunctions? 
\item What is their normalization?
\item What distinguishes equilibrium from nonequilibrium processes in terms of fluctuations, dynamical or otherwise?
\end{itemize}

The fact that we have to deal with non-Hermitian operators is the main reason why studying nonequilibrium processes is more challenging than studying equilibrium processes, but at the same time so much more interesting. We are far from completely understanding nonequilibrium systems, and I do not believe, in fact, that there is or will be a general theory of these systems (not at least in the way many physicists picture it). But I do believe that having a good mathematical grounding in the subject -- at least as good as the one we expect in quantum mechanics or any modern physics subject -- is necessary for making progress.

Obviously, such a pedagogical goal cannot be achieved in 20-odd pages of notes (a book is in progress), so I decided to focus here on Langevin-type equations modelling noisy systems driven by external forces and baths, sacrificing mathematical rigor, as always, for clarity. With this, one should keep in mind that all  the results discussed can be applied more generally (sometimes with minor modifications) to a large class of Markov processes, including Markov chains and Markov jump processes. Some references and exercises on those are given in the text.

\section{Markov processes}

\subsection{Stochastic differential equations}

We consider throughout these notes Markov processes defined by the following \myemph{stochastic differential equation (SDE)}:
\be
dX_t = F(X_t) dt+\sigma dW_t,
\label{eqsde1}
\ee
where
\begin{itemize}
\setlength\itemsep{0.1em}
\item $X_t$ is a vector in $\reals^n$ representing the \myemph{state} of the system at the time $t$. Note that we do not use bold symbols to represent vectors.
\item $F:\reals^n\ra\reals^n$ is a vector field that drives the deterministic evolution of $X_t$ when there is no noise ($\sigma=0$). We call this function the \myemph{force} or the \myemph{drift} of the system. It can explicitly depends on time, for example, by having $F(X_t,t)$, but we do not consider this possibility here.

\item $W_t$ is a vector of independent \myemph{Brownian} or \myemph{Wiener motions}, whose increments $dW_t$ are Gaussian-distributed with zero mean and variance $dt$. In many applications, $W_t$ is taken to have as many components as $X_t$, so that $W_t\in\reals^n$, but it is also possible to have $W_t\in\reals^m$ with $m>n$ or $m<n$. For simplicity, we consider here $m=n$.

\item $\sigma$ is the \myemph{noise matrix}, assumed here not to depend on $X_t$ to simplify the discussion. That matrix has dimensions $n\times m$ to match the dimensions of $X_t$ and $W_t$. For $W_t\in \reals^n$, it is an $n\times n$ matrix.
\end{itemize}

The SDE above governing the evolution of $X_t$ (as an infinitesimal difference equation) is called in mathematics an \myemph{It\^o SDE}. For the purpose of these notes, we can consider $X_t$ to be defined equivalently by a noisy \myemph{ordinary differential equation (ODE)} having the more common form
\be
\dot X_t = F(X_t) +\sigma \xi_t,
\ee
where $\xi_t$ is an $n$-dimensional \myemph{Gaussian white noise} corresponding formally to the time-derivative of $W_t$ and defined by the properties
\be
\langle \xi_t\rangle=0,\qquad \langle \xi^i_t\xi^j_{t'}\rangle = \delta_{ij} \delta(t-t'),
\ee
where $i$ and $j$ denote specific components of $\xi(t)$ and $\langle\cdot\rangle$ denotes the expectation or mean (with respect to whatever random variable or process found in the brackets).

I assume in these notes that the reader knows that Brownian motion is nowhere differentiable, which explains why it is preferable to express a diffusion process in the It\^o rather than in the noisy ODE form. I also assume some familiarity with stochastic calculus, though this is not essential, as long as one is aware that the rules for calculating derivatives and integrals of stochastic processes differ slightly from those of normal calculus. For background references on SDEs and stochastic calculus, see the Further reading section. Some important differences between stochastic and normal calculus will be pointed out along the notes.

\subsection{State distribution and generator}

The first natural problem to consider when studying a Markov process is to determine its probability density $P(X_t=x)=p(x,t)$, starting from an initial density $p(x,0)$ that represents the distribution with which we sample the initial state $X_0$. For the general SDE shown in (\ref{eqsde1}), that density is known to be given by the \myemph{Fokker-Planck equation},
\be
\p_t p(x,t) = -\nabla \cdot (F(x) p(x,t))+\frac{1}{2}\nabla\cdot D\nabla p(x,t),
\label{eqfp1}
\ee
which involves the symmetric matrix $D=\sigma\sigma^T$, called the \myemph{covariance matrix}. To emphasize that this partial differential equation (PDE) is linear, we can express it as
\be
\p_t p(x,t) = L^\dag p(x,t),
\label{eqfp2}
\ee
where 
\be
L^\dag = -\nabla\cdot F +\frac{1}{2}\nabla\cdot D\nabla
\ee
is a linear differential operator called the \myemph{Fokker-Planck generator}. The analogy with the Schr\"odinger equation of quantum mechanics should be obvious.

The Fokker-Planck equation can also be rewritten, as is well known, as a conservative equation involving the (vector) probability current
\be
J_t(x)=F(x)p(x,t)-\frac{D}{2}\nabla p(x,t),
\ee
called the \myemph{Fokker-Planck current}, as
\be
\p_t p(x,t) +\nabla\cdot J_t(x) =0.
\ee
This form is useful for interpreting the nature of the \myemph{stationary probability density} $p_s(x)$ satisfying
\be
L^\dag p_s=0 
\ee
or equivalently $\nabla\cdot J_s=0$, where $J_s$ is the \myemph{stationary current} associated with $p_s$. This shows that $p_s(x)$ is the eigenfunction of $L^\dag$ with eigenvalue $0$. When the process $X_t$ is \myemph{ergodic}, $p(x,t)\ra p_s(x)$ from any initial density as $t\ra\infty$.

For the next sections, it is useful to note that the Fokker-Planck equation also determines the evolution of expectations of $X_t$ having the general form
\be
\langle f(X_t)\rangle =\int p(x,t) f(x)dx,
\ee
where $f$ is any smooth function of the process. Indeed, it can be proved using the natural inner product defined by the expectation  that
\be
\p_t \langle f(X_t)\rangle =\langle (Lf)(X_t)\rangle,
\label{eqev1}
\ee
where $L$ is the adjoint of $L^\dag$ (see Appendix~\ref{appspace}). This operator, which is simply called the \myemph{generator} of $X_t$, acts on the function $f$ rather than the density $p$, as is explicit from the notation above, and is equal here to
\be
L= F\cdot \nabla+\frac{1}{2}\nabla\cdot D\nabla.
\ee

This result is much less known in physics than the Fokker-Planck equation, though it is as important. It corresponds, in a way, to the Heisenberg picture of quantum mechanics which describes the evolution of observables (after taking their expectation), as opposed to the Schr\"odinger picture which describes the evolution of probabilities. The generator will become useful for treating large deviations.

\subsection{Examples}

There are many standard SDEs used in physics to model noisy systems driven by forces, external reservoirs, heat baths and noise sources in general. The following is a representative list -- for more examples, see the reading list in Sec.~\ref{secread1}.
\begin{itemize}
\setlength\itemsep{0.1em}
\item \myemph{Kramers or underdamped Langevin equation:}
\begin{eqnarray}
d q_t &=& \frac{p_t}{m}d t\nonumber\\
d p_t &=& \left(-\nabla V(q_t) +\phi_t-\Gamma \frac{p_t}{m}\right)d t+\sqrt{2\Gamma/\beta}\, d W_t,
\label{eqkram1}
\end{eqnarray}
where $q_t$ is the position, $p_t$ the momentum, $V(q)$ is a potential, $\Gamma$ is the friction, $\phi_t$ is an external force, and $\beta$ is the inverse temperature of the thermal noise. The noise acts as a force in Newton's equation and so only affects the momentum, not the position.

\item \myemph{Overdamped Langevin equation:} 
\be
d q_t=\Gamma^{-1} (-\nabla V+\phi_t)d t+\sqrt{2\Gamma^{-1}/\beta}\, d W_t.
\label{eqodeq1}
\ee
This is an SDE for the position obtained by taking the overdamped ($m\ra 0$) limit of Kramers equation.

\item \myemph{Gradient SDEs:}
\be
dX_t = -\nabla U(X_t) dt+\sigma dW_t
\label{eqgradsde1}
\ee
with $\sigma$ proportional to the identity matrix, that is, $\sigma =\eps\id$. The stationary density of this SDE is the Gibbs distribution
\be
p_s(x)=c\, e^{-2U(x)/\eps^2},
\label{eqgibbs1}
\ee
where $c$ is a normalization constant (see Exercise~\ref{ex3}).

\item \myemph{Linear diffusions:}
\be
dX_t = -MX_t dt+\sigma dW_t,
\label{eqsdelin1}
\ee
where $M$ is an $n\times n$ matrix assumed to be positive definite (positive eigenvalues) in order for $X_t$ to have a stationary density (see Exercise~\ref{ex4}).

\item \myemph{Ornstein-Uhlenbeck process:}
\be
dX_t =-\gamma X_t dt+\sigma dW_t
\label{eqou1}
\ee
with $X_t\in\reals$ and $W_t\in\reals$. This is obviously a gradient SDE with quadratic potential $U(x)=\gamma x^2/2$ having a Gibbs stationary distribution with $\sigma=\eps$.
\end{itemize}

\subsection{Equilibrium versus nonequilibrium processes}

The distinction between equilibrium and nonequilibrium systems in the context of stochastic processes is based on the notion of time reversibility or, equivalently, detailed balance. It would take too much space to fully explain these notions, so we only summarize them. The idea, essentially, is that a process is an \myemph{equilibrium process} if the probability of any given trajectory is the same as the probability of that trajectory reversed in time. If that is not the case, then the process is \myemph{nonequilibrium}. 

For Markov processes, it can be proved that this definition of equilibrium in terms of ``forward'' and ``backward'' trajectories is equivalent to the notion of detailed balance, which is itself related (in most cases) to having a vanishing probability current in the Fokker-Planck equation (see Exercise~\ref{ex2}). Moreover, all of these notions are related in general to the eigenvalues of the generator $L$. 

We summarize these connections in Table~\ref{table2}, assuming that the process $X_t$ is stationary, that is to say, it has a stationary distribution and its initial condition $X_0$ is drawn according to that distribution.

\begin{table}[t]
\centering
\begin{tabular}{cccc}
\toprule
\myemph{Equilibrium} 		& & & \myemph{Nonequilibrium}\\
\midrule
Reversible			& & & Non-reversible\\
$J_s(x)=0$ 			& & & $J_s(x)\neq 0$ although $\nabla\cdot J_s=0$\\
Spectrum of $L$ real		& & & Spectrum of $L$ generally complex\\
\bottomrule
\end{tabular}
\caption{Comparison of equilibrium and nonequilibrium Markov processes in steady states.}
\label{table2}
\end{table}

\subsection{Comparison with quantum mechanics}

It is useful at this point to reflect on the structure of Markov diffusions, especially the linear form of the Fokker-Planck equation and its generator, by comparing it to what we know about the evolution of quantum systems -- see Table~\ref{tabqmcomp}. 

From this table, you should see that the theory of Markov processes has much in common with quantum mechanics, as announced in the introduction. Both are  linear theories of evolution for a ``vector'' corresponding to the probability density $p(x,t)$ for Markov processes and the wavefunction $\psi(x,t)$ for quantum systems. As a result, the central object in both theories is the generator of that evolution, which is the Fokker-Planck operator $L^\dag$ for Markov processes and the Hamiltonian $H$ for quantum systems. 

The main difference between the two sides is that, unlike the Hamiltonian of closed quantum systems, the generator of Markov processes is not always self-adjoint, a property deeply related to whether we are dealing with an equilibrium or nonequilibrium process. This means, in practice, that we have to be careful when diagonalizing matrices or operators and dealing, in general, with eigenvectors or eigenfunctions (see Appendix~\ref{appnh}). 

\begin{table}[t]
\centering
\begin{tabular}{ccccc}
\toprule
			& & \myemph{Markov} 		& & \myemph{Quantum} \\
\midrule
State			& & $X_t$ 			& & $|\psi(t)\rangle$ or $\psi(x,t)$ \\
Distribution 	& & $p(x,t)$ 			& & $|\psi(x,t)|^2$\\
Evolution 	 	& & Fokker-Planck		& & Schr\"odinger\\
Generator 	& & $L$				& & $H$ (Hamiltonian)\\
Propagator	& & $U(t) = e^{L^\dag t}$		& & $U(t)=e^{-iHt/\hbar}$\\
Inner product	& & $\langle p,f\rangle$ 	& & $\langle \psi,\psi\rangle=\langle \psi|\psi\rangle$ \\
Duality		& & $\langle p,Af\rangle=\langle A^\dag p,f\rangle$ & & $\langle \psi, A\psi\rangle=\langle A\psi, \psi\rangle=\langle\psi|A|\psi\rangle$\\
Self-adjoint?	& & Not necessarily		& & Always\\
\bottomrule
\end{tabular}
\caption{Comparison between Markov processes and quantum mechanics.}
\label{tabqmcomp}
\end{table}

\subsection{Further reading}
\label{secread1}

\begin{itemize}
\setlength\itemsep{0.1em}
\item SDEs and Langevin equations: \cite{jacobs2010}. For more technical yet readable presentations, see \cite{brzezniak1999} and especially \cite{pavliotis2014}.

\item Stochastic calculus: \cite{jacobs2010,oksendal2000} for the theory and \cite{higham2001} for simulations.

\item Theory of Markov processes focusing more on Markov chains and jump processes (also called continuous-time Markov chains): A very good reference, for its scope, clarity and number of exercises, is \cite{grimmett2001}. Though a maths textbook, it should be compulsory reading for any serious students in statistical physics. For a leaner textbook, see \cite{brzezniak1999}.

\item Fokker-Planck equation: \cite{risken1996}.

\item Stochastic processes with applications in physics: \cite{gardiner1985,kampen1992}. In my opinion, statistical physics is in real need of a modern reference on that front.
\end{itemize}

\section{Dynamical observables}

The study of Markov processes in physics has focused a lot in the past on the statistical properties of the state $X_t$ -- its distribution, its average, in addition to its variance and covariance, which can be related to diffusion and transport coefficients \cite{reichl1980}. In the last 20 years or so, researchers have also become interested in the statistics of \textit{time-integrated quantities involving} $X_t$. An example is the \myemph{mechanical work} done by the force $F$ on $X_t$ over a time interval $[0,T]$, as calculated by
\be
W_T=\int_0^T F(X_t) \circ dX_t,
\ee   
where the circle $\circ$ indicates that the integral is to be calculated using the midpoint Riemann integral rule, also called the \myemph{Stratonovich convention}.\footnote{The work is a scalar quantity, so the product $F(X_t)\circ dX_t$ is also a scalar product.} This random variable depends obviously not only on the state $X_t$ at time $t$, but on the whole trajectory of this process between $t=0$ and $t=T$. Thus, for different (random) trajectories, one typically gets different (random) work values. For this reason, $W_T$ is often called an \myemph{additive functional} of the process or, more physically, a \myemph{dynamical observable}. 

We list next other physical examples:
\begin{itemize}
\setlength\itemsep{0.1em}
\item \myemph{Potential energy:} The change in time of potential energy can be written as
\be
\Delta U_T = U(X_T)-U(X_0) = \int_0^T \nabla U(X_t)\circ dX_t.
\ee
This holds whether the SDE of $X_t$ is gradient or not, but only if the integral is interpreted with the Stratonovich convention, which follows the standard rules of calculus. In the It\^o convention, corresponding to the left-point Riemann rule, there would be additional terms in the integral, coming from It\^o's calculus \cite{jacobs2010}, which are unphysical. This explains why the work $W_T$ must be defined in the Stratonovich convention: for a potential force, the work is the change of potential.

\item \myemph{Empirical distribution:}
\be
\rho_T(x) = \frac{1}{T}\int_0^T \delta (X_t-x)\, dt.
\label{eqed1}
\ee
This random function represents the fraction of time in $[0,T]$ that the process $X_t$ ``takes'' the value $x$. For an ergodic process, it converges to the stationary density $p_s(x)$ for almost all trajectories. Mathematically, we say that $\rho_T$ converges in probability to $p_s$ as $T\ra\infty$.

\item \myemph{Empirical current:} 
\be
J_T(x)=\frac{1}{T}\int_0^T \delta(X_t-x)\circ dX_t.
\label{eqec1}
\ee
This random field represents intuitively the local mean ``velocity'' of $X_t$ at $x$, if we view $dX_t$ formally as $\dot X_t dt$. It is more aptly called the empirical current because it represents a flow at each point $x$ that converges in probability to the stationary Fokker-Planck current $J_s(x)$ as $T\ra\infty$ (see Exercise~\ref{ex7}). The Stratonovich rule is also important for this convergence.

\item \myemph{Entropy production:}
\be
\Sigma_T = 2\int_0^T (D^{-1}F(X_t))\circ dX_t.
\label{eqep1}
\ee
This is an important quantity in the theory of nonequilibrium systems, related to the non-reversibility (or nonequilibrium nature) of $X_t$ and, as is clear from its definition, also to the work $W_T$ \cite{lebowitz1999}. The factor 2 is there to obtain the normal thermodynamic relation $\Sigma_T = \beta \Delta V_T$ for the overdamped Langevin equation (\ref{eqodeq1}) (check this).
\end{itemize}

From these examples, it is natural to consider a general class of dynamical observables having the form
\be
A_T = \frac{1}{T}\int_0^T f(X_t) dt+\frac{1}{T}\int_0^T g(X_t)\circ dX_t,
\label{eqobs1}
\ee
where $f$ (scalar) and $g$ (vector) are two arbitrary functions that depend on the system and physical quantity considered, and $\circ$ denotes, as before, the Stratonovich (scalar) product. This choice of convention is actually not important -- we could use the It\^o convention instead with a slight modification of the results that will come next. What is more important is the factor $1/T$ which is there for two reasons: first, to make $A_T$ intensive in time and, second, to guarantee, following the examples of the empirical density and empirical current, that $A_T$ converges in probability to a constant (its mean) in the long-time limit, $T\ra\infty$. 
 
\section{Large deviations}

\subsection{Large deviation principle}

Our goal now is to study the probability distribution $P(A_T=a)$ of a given dynamical observable $A_T$ and Markov process $X_t$. In general, it is very difficult to obtain that distribution exactly. In many cases, however, it is known that it has the following general asymptotic form:
\be
P(A_T=a)\approx e^{-TI(a)}
\label{eqldp1}
\ee
in the limit of large observation time $T$. The meaning of this approximation, which is called the \myemph{large deviation principle (LDP)}, is that the dominant contribution of $P(A_T=a)$ is a decaying exponential, and so that any corrections to that contribution is sub-exponential in $T$.\footnote{See Appendix~B of \cite{touchette2009} or Sec.~1.2 of \cite{dembo1998} for the mathematical definition of the LDP. The loose definition given here assumes that $P(A_T=a)$ is a probability density.} The exponent or \myemph{rate function} $I(a)$ is therefore the essential information that we need to find in order to characterize the fluctuations of $A_T$. In particular,
\begin{itemize}
\item $I(a)\geq 0$ for all $a$, so $P(A_T=a)$ decays exponentially fast with $T$, except for values $a$ such that $I(a)=0$. 
\item For Markov processes, there is usually only one point $a^*$ where $I(a^*)=0$, so if $P(A_T=a)$ decays where $I(a)>0$, it must concentrate on $a^*$ by conservation of probability.
\item The zero of $I(a)$ is a reflection of the Law of Large Numbers: it gives the most probable or \myemph{typical value} of $A_T$ in the long-time limit and coincides with the mean of $A_T$. In an experiment where the work $W_T$ per unit time would be measured, for example, one would see that most trajectories do work close to its mean value $\langle W_T\rangle$. Only rarely would we see trajectories that require more or less work than this average.
 
\item The rate function $I(a)$ characterizes the rare fluctuations of $A_T$ around this typical value. In general, it is not a parabola, so fluctuations of dynamical observables are in general not Gaussian. This is the most important point about large deviation theory~--~the fact that it characterizes fluctuations beyond Gaussian fluctuations and so beyond the Central Limit Theorem.
\end{itemize}

These properties will become clear once we start calculating rate functions for specific processes and observables. 

We describe next the most common way of obtaining rate functions using a result of large deviation theory known as the G\"artner-Ellis Theorem, which essentially boils down, for additive functionals of Markov processes, to calculating a dominant eigenvalue of a linear operator. This is not an easy task to carry out in many applications, but it is definitively simpler than calculating the exact distribution of $A_T$.

\subsection{Spectral problem}
\label{secge}

The \myemph{G\"artner-Ellis Theorem} is based on the following function:
\be
\lambda(k)=\lim_{T\ra\infty}\frac{1}{T}\ln \langle e^{TkA_T}\rangle,
\label{eqscgf1}
\ee
known as the \myemph{scaled cumulant generating function (SCGF)}. The theorem says, in its simplified version, that if $\lambda(k)$ exists for $k\in\reals$ and is differentiable in $k$, then
\begin{enumerate}[1)]
\item $A_T$ satisfies a large deviation principle, so its distribution has the scaling form (\ref{eqldp1});
\item Its rate function $I(a)$ is the \myemph{Legendre-Fenchel transform} of $\lambda(k)$:
\be
I(a) = \max_{k\in\reals} \{ka-\lambda(k)\}.
\label{eqlf1}
\ee
\end{enumerate}

The question now is, how do we calculate $\lambda(k)$? 

For this, we can use another result of probability theory, known as the Feynman-Kac formula, to cast the evolution of the generating function of $A_T$ as a linear PDE similar to the Fokker-Planck equation, which involves some linear operator $\cL_k$, called the \myemph{tilted generator}, and then study the asymptotic evolution of that PDE to realize that it is dominated by the dominated eigenvalue of $\cL_k$. These steps are presented in Appendix~\ref{appfk} and lead to the main result of these notes, namely,
\be
\lambda(k)=\zeta_{\max}(\cL_k),
\ee
where $\zeta_{\max}(\cL_k)$ denotes the dominant eigenvalue of $\cL_k$. To be more precise, \textit{$\lambda(k)$ is equal to $\zeta_{\max}(\cL_k)$ whenever $\lambda(k)$ exists as a SCGF}. This is an important precision (see Exercise~\ref{ex14b}). For the SDE (\ref{eqsde1}) and the observable $A_T$ defined in (\ref{eqobs1}), the tilted generator is explicitly given by
\be
\cL_k = F\cdot (\nabla +kg)+\frac{1}{2}(\nabla+kg)\cdot D(\nabla+kg)+kf,
\label{eqlk1}
\ee
where $f$ and $g$ are the functions entering in $A_T$ (see Appendix~\ref{appfk} and Exercise~\ref{ex8}). Note that $\cL_{k=0}=L$ and so $\lambda(0)=\zeta_{\max}(L)=0$.

This result is a PDE generalization of the Perron-Frobenius Theorem about positive matrices, guaranteeing that $\lambda(k)$ is real, and applies essentially whenever the spectrum $\cL_k$ is gapped. The resulting dominant eigenvalue problem is similar to a quantum eigenvalue problem, except that $\cL_k$ is not in general a Hermitian operator. This makes the calculation of $\lambda(k)$ more complicated. 

In fact, compared to quantum mechanics, we must now not only consider the eigenvalue problem
\be
\cL_k r_k(x) = \lambda(k) r_k(x),
\ee
where $r_k(x)$ is the ``right'' eigenfunction associated with the dominant eigenvalue $\lambda(k)$. We must also solve in parallel the dual eigenvalue problem
\be
\cL_k^\dag l_k(x) = \lambda(k) l_k(x),
\ee
where $l_k(x)$ is the corresponding ``left'' eigenfunction, by requiring overall the following boundary condition:
\be
r_k(x)l_k(x)\overset{|x|\ra \infty}{\longrightarrow} 0.
\label{eqbc1}
\ee
The reason for this, in short, is that the duality between $\cL_k$ and $\cL_k^\dag$ is equivalent to performing integration by parts (Appendix~\ref{appspace}) and $r_k(x)l_k(x)$ is the boundary term in that integration that must vanish at infinity. Therefore, we see that, contrary to quantum mechanics, we cannot just solve the direct eigenvalue problem for $\cL_k$ by requiring that $r_k(x)$ alone decays to 0 at infinity -- we must consider the direct and dual eigenvalue problems  at the same time with (\ref{eqbc1}) to find the correct dominant eigenvalue.

This duality also determines the correct normalization to use for the eigenfunctions, which turns out to be
\be
\int r_k(x)\, l_k(x)\, dx=1
\label{eqnorm1}
\ee  
(see Appendix~\ref{appnh}). For convenience, we also impose
\be
\int l_k(x)\, dx=1.
\label{eqnorm2}
\ee
Note that $l_{k=0}=p_s$ and $r_{k=0}=1$ (see Exercise~\ref{ex8b}), so both normalization conditions reduce to $\int p_s(x) dx=1$ for $k=0$.

\subsection{Symmetrization}

The tilted generator $\cL_k$ is known, in many cases, to have a real spectrum even though it is not Hermitian.\footnote{Being Hermitian is only a sufficient condition for having a real spectrum, not a necessary condition.} This happens, for example, when dealing with gradient SDEs and observables such that $g=0$. In this case, it should be possible to transform $\cL_k$ to a Hermitian operator $\cH_k$ in an unitary way (so as to preserve the spectrum) and then work only with $\cH_k$ using techniques from quantum mechanics to find $\lambda(k)$.

For the case of gradient SDEs mentioned, this is indeed possible and the unitary transformation or \myemph{symmetrization} that we must use is given by
\be
\cH_k = p_s^{1/2} \cL_k p_s^{-1/2},
\label{eqsym1}
\ee
where $p_s(x)$ is the Gibbs stationary distribution of the SDE shown in Eq.~(\ref{eqgibbs1}) (see also Exercise~\ref{ex3}). Note that this is an operator transformation: when applied to a function $\phi$, $\cH_k$ acts as follows:
\be
\cH_k\phi = p_s^{1/2} (\cL_k p_s^{-1/2} \phi),
\ee
so that $\cL_k$ is applied to the product $p_s^{-1/2}\phi$. The resulting function is then multiplied by $p_s^{1/2}$. 

For a gradient SDE defined in (\ref{eqgradsde1}) and an observable $A_T$ such that $f\neq 0$ but $g=0$, it is not difficult to see by direct calculation (see Exercise~\ref{ex10}) that $\cH_k$ has the form
\be
\cH_k = \frac{\eps^2}{2}\Delta-V_k,
\label{eqh1}
\ee
where $\Delta = \nabla^2$ is the Laplacian, $\eps$ is the noise amplitude of the SDE, and
\be
V_k(x) = \frac{|\nabla U(x)|^2}{2\eps^2}-\frac{\Delta U(x)}{2}-kf(x)
\label{eqv1}
\ee
is an effective quantum-like potential. Our eigenvalue problem thus reduces to 
\be
\cH_k \psi_k = \lambda(k) \psi_k,
\ee
where $\psi_k$ is the eigenfunction of $\cH_k$ associated with $\lambda(k)$. 

We recognize in this equation the time-independent Schr\"odinger equation, up to a minus sign, with $\eps^2 =\hbar^2/m$ and potential $V_k$. The sign difference means that what we find as the dominant (largest) eigenvalue of $\cH_k$ corresponds to the ground state (smallest) energy of $-\cH_k$.

The eigenvalue $\lambda(k)$ is the same for $\cL_k$ and $\cH_k$ since these two operators are unitarily related. The eigenfunctions $\psi_k$ of $\cH_k$, however, are different from the eigenfunctions $r_k$ of $\cL_k$. In general, the two are related together (see Exercise~\ref{ex10b})  by 
\be
\psi_k(x) = p_s(x)^{1/2} r_k(x).
\label{eqt1}
\ee
Moreover, it can be verified that
\be
\psi_k(x) = p_s(x)^{-1/2}l_k(x).
\label{eqt2}
\ee
This is interesting because it implies that the boundary condition (\ref{eqbc1}) that we have for the full eigenvalue problem now reduces to
\be
\psi(x)^2\overset{|x|\ra \infty}{\longrightarrow} 0,
\ee
which is the normal quantum boundary condition (modulo a complex conjugate), while the normalization condition in (\ref{eqnorm1}) reduces to
\be
\int \psi(x)^2\, dx =1.
\ee
In a more practical way, it also implies that we can now focus on only one eigenvalue problem with natural (quantum) boundary conditions imposed only on $\psi_k$. 

This symmetrization method greatly simplifies, obviously, the calculation of large deviation functions. A natural question is, to which class of Markov processes can it be applied to besides gradient SDEs? Table~\ref{table3} gives some answers. The basic idea is that, if we consider an equilibrium process and an ``equilibrium-type'' observable characterized by $g=0$ or $g$ gradient, then $\cL_k$ can be symmetrized. If $X_t$ is a nonequilibrium process, then $\cL_k$ cannot be symmetrized in general because $L$ itself cannot be symmetrized. Finally, if $X_t$ is reversible but $g$ is not gradient, then $\cL_k$ generally cannot be symmetrized because the fluctuations of the observable that we consider are essentially created in a nonequilibrium way (see Exercise~\ref{ex12}).

\begin{table}[t]
\centering
\begin{tabular}{ccccccc}
\toprule
	$X_t$				& &	$L$ symmetrizable?		& & $g$	 		& & $\cL_k$ symmetrizable?\\
\midrule
	Reversible	 		& & 	Yes					& & 0			& & Yes\\
						& & 						& & Gradient $g=\nabla\varphi$		& & Yes\\
						& &						& & Non-gradient	& & No \\
	\hline
	Non-reversible		 	& & 	No (generally)					& & Any			& & No (generally) \\
\bottomrule
\end{tabular}
\caption{Spectral problem for equilibrium and nonequilibrium processes.}
\label{table3}
\end{table}

\subsection{Example: Ornstein-Uhlenbeck process}
\label{secexou}

We close these notes by applying the material of the previous sections to a specific example involving the Ornstein-Uhlenbeck process defined in~(\ref{eqou1}). Our goal is to find the rate function $I(a)$ characterizing the fluctuations of the following observable:
\be
A_T=\frac{1}{T}\int_0^T X_t\, dt,
\label{eqlinobs1}
\ee
which represents mathematically the area under the paths of the process, and which can be related physically to the mechanical work performed by a laser tweezer on a Brownian particle immersed in water \cite{zon2003a}. 

To obtain the rate function, we first write down the tilted generator associated with this process and observable, noting in this case that $f(x)=x$ and $g=0$:
\be
\cL_k = L+kf = -\gamma x\frac{d}{dx}+\frac{\eps^2}{2}\frac{d^2}{dx^2}+k x,\qquad k\in\reals.
\ee
This operator is not Hermitian, but since $X_t$ is gradient (any SDE on $\reals$ is gradient) and $g=0$, it can be symmetrized with its (Gaussian) Gibbs distribution
\be
p_s(x)=\sqrt{\frac{\gamma}{\pi \eps^2}}\, e^{-\gamma x^2/\eps^2}\propto e^{-2U(x)/\eps^2}
\label{eqous1}
\ee 
to
\be
\cH_k=\frac{\eps^2}{2}\frac{d^2}{dx^2}-\frac{\gamma^2 x^2}{2\eps^2}+\frac{\gamma}{2}+kx.
\ee

We recognize the Hamiltonian of the 1D quantum harmonic oscillator if we multiply by $-1$ and shift the space to $x\ra x+\eps^2k/\gamma^2$. The SCGF, corresponding to minus the known ground state of the quantum oscillator which does not depend on the shift (see any textbook on quantum mechanics), is therefore found to be
\be
\lambda(k)=\frac{\eps^2 k^2}{2\gamma^2}.
\ee
This exists and is differentiable for all $k\in\reals$, so we can the apply the Legendre-Fenchel transform (\ref{eqlf1}), which in this case reduces to a simple Legendre transform (why?), to finally obtain
\be
I(a)=\frac{\gamma^2 a^2}{2\eps^2}.
\ee
This shows that the fluctuations of $A_T$ are Gaussian around the typical value $A_T=0$, a result consistent with the fact that linear integrals of Gaussian processes are Gaussian.

Although we do not need the eigenvectors $r_k(x)$ and $l_k(x)$ in the calculation of the SCGF and rate function, it is instructive to compute them. For the quantum oscillator, it is well known that the eigenfunctions are given in terms of Hermite polynomials. Considering only the ground state, we find here
\be
\psi_k(x) =\left(\frac{\gamma }{\pi \eps^2}\right)^{1/4}\, \exp\left({-\frac{\gamma \left(x-\eps^2k/\gamma^2\right)^2}{2\eps^2}}\right).
\ee
Using the transformations (\ref{eqt1}) and (\ref{eqt2}), and applying the normalization conditions (\ref{eqnorm1}) and (\ref{eqnorm2}), we then find
\be
r_k(x)=\exp\left(\frac{k x}{\gamma }-\frac{3 \eps^2 k^2}{4 \gamma ^3}\right)
\ee
and
\be
l_k(x)= \sqrt{\frac{\gamma}{\pi\eps^2}}\, \exp\left(-\frac{\gamma\left(2 x-\eps^2k/\gamma^2\right)^2}{4 \eps^2}\right).
\ee
This clearly shows that $r_k(x)$ or $l_k(x)$ do not decay to zero as $x\ra\pm\infty$~--~it is their product again that does so, in agreement with (\ref{eqbc1}). Finally, note that $r_{k=0}(x)=1$ while $l_{k=0}(x)=p_s(x)$, as pointed out before (see also Exercise~\ref{ex8b}).

The exercises found at the end of these notes extend this simple calculation to other equilibrium-type observables, before slowly moving into the realm of nonequilibrium processes. For more information about nonequilibrium large deviations, and the current research done on this topic, see the pointer references next.

\subsection{Further reading}

\begin{itemize}
\setlength\itemsep{0.1em}

\item Mathematical theory of large deviations: \cite{dembo1998,hollander2000}.
\item Applications of large deviations in statistical physics: \cite{touchette2009,harris2013}.
\item Quantum approach to large deviations: \cite{majumdar2002,majumdar2005}.
\item Related approach based on the so-called thermodynamics of trajectories: \cite{lecomte2005,lecomte2007,lecomte2007c}.
\item Stochastic thermodynamics: \cite{seifert2012}.
\item Fluctuations of interacting particle systems modelling particle and energy transport: \cite{mallick2015,derrida2007,bertini2007}.
\item Fluctuation relations and symmetries: \cite{harris2007,lebowitz1999}.
\item Recent theory of fluctuation processes explaining, with modified SDEs, how large deviations are created in time: \cite{chetrite2013,chetrite2014,chetrite2015}.
\item Entropy production: \cite{lebowitz1999}.
\item Fluctuations of empirical density and current (level 2.5 of large deviations): \cite{barato2015}.
\item Large deviations for open quantum systems: \cite{garrahan2010} and Garrahan's lecture notes.
\item Low-noise large deviations: \cite{touchette2009}.
\item Large deviation simulations: \cite{bucklew2004,touchette2011}.
\end{itemize}

\appendix

\section{Dual spaces for Markov processes}
\label{appspace}

The expectation
\be
\langle f(X_t)\rangle = \int p(x,t) f(x)\, dx
\ee 
of a function $f$ of $X_t$ defines the following natural \myemph{scalar} or \myemph{inner product} in the theory of Markov processes:
\be
\langle p, f\rangle=\int p(x) f(x) \, d x,
\ee
which connects the space of normalized probability densities and the space of functions of $X_t$, also called \myemph{test functions} or \myemph{observables}. (See Table~\ref{tabqmcomp} for a comparison of this inner product and the one used in quantum mechanics.)

Applying an operator either on $f$ or on $p$ leads us to define the notion of \myemph{dual} or \myemph{adjoint operator}: if $L$ acts on $f$, then its adjoint $L^\dag$ acts on $p$ according to
\be
\langle p,Lf\rangle= \langle L^\dag p,f\rangle.
\ee
Because the integral defining the inner product is performed with $dx$, the Lebesgue measure, we say that $L^\dag$ is the adjoint of $L$ with respect to the Lebesgue measure.

For differential operators, the duality between $L$ and $L^\dag$ simply corresponds to performing integration by parts:
\be
\langle p,\frac{df}{dx}\rangle =\int p df = \left. pf\right|_{\text{boundary}} - \int f dp,
\ee
which leads to $(\nabla)^\dag = -\nabla\cdot$ for the gradient and $(\Delta)^\dag =\Delta$ for the Laplacian, if we choose $p$ and $f$ such that the boundary term (usually at infinity) vanishes. The Fokker-Planck equation also follows from this duality by noting from (\ref{eqev1}) that
\be
\p_t\langle f(X_t)\rangle= \langle p_t,Lf\rangle = \langle L^\dag p_t,f\rangle.
\ee
Since this holds for any test functions, we must then have (\ref{eqfp1}).

\section{Non-Hermitian operators}
\label{appnh}

Non-Hermitian or non-self-adjoint operators (we do not make a difference between the two here) can be diagonalized in the same way as Hermitian operators in quantum mechanics~--~just think about symmetric versus non-symmetric matrices. The only difference is that, being non-Hermitian (essentially, non-symmetric), we have to distinguish between ``left'' and ``right''  eigenfunctions (or eigenvectors) to build their spectral decomposition.

To be precise, let us consider a linear differential operator $L$ and consider the eigenvalue problem
\be
Lv(x)=\lambda v(x),
\ee
which can be solved for a set of eigenvalues $\lambda$, called the \myemph{spectrum} of $L$, and their corresponding eigenfunctions $v(x)$. We do not put indices to these objects as is commonly done -- their numbering or labelling is implicit. If $L$ is not self-adjoint, the dual eigenvalue problem
\be
L^\dag u(x)= \beta u(x)
\label{eqdep1}
\ee
has a spectrum that is the complex conjugate of the spectrum of $L$, that is, $\lambda=\beta^*$, but will in general have a completely different set of eigenfunctions $u(x)$.

As in quantum mechanics, eigenfunctions $u$ or $v$ associated with distinct eigenvalues are orthogonal, so that
\be
\langle u_i,v_j\rangle= \int u_i^*(x) v_j(x) \, dx=\delta_{ij}
\ee
by properly normalizing them. Moreover, it can be shown that
\be
\delta(x-x')=\sum_i u_i^*(x) v_i(x'),
\label{eqcomp1}
\ee
a property known in quantum mechanics as the \myemph{completeness relation}. As a result, we see that the set of eigenfunctions $u$ and $v$ form a complete basis onto which any function can be decomposed. For an application of this decomposition for solving the Fokker-Planck equation, see Sec.~5.4 of \cite{risken1996}; for its application to our problem of finding large deviation functions, see the next appendix on the Feynman-Kac formula.

If $L$ is self-adjoint, then $u=v$ and we recover the usual complete basis of quantum mechanics. Moreover, if $L$ is a matrix, then the dual eigenvalue problem (\ref{eqdep1}) is equivalent to
\be
u^\dag L =\beta^\dag u^\dag = \lambda u^\dag,
\ee
where $u^\dag$ is now seen as a row vector multiplying $L$. In this sense, it is common to call $u$ the \myemph{left eigenvector} of $L$ and $v$ the \myemph{right eigenvector} of $L$. 


\section{Feynman-Kac formula}
\label{appfk}

The linear structure of the Fokker-Planck equation (\ref{eqfp2}) means that we can write down the time-dependent density $p(x,t)$ as
\be
p(x,t)=U(t)p(x,0),
\ee
where $U(t)=e^{tL^\dag}$ is an operator acting on the initial density, called the \myemph{propagator}.\footnote{This applies to homogeneous SDEs with time-independent drift. For a time-dependent drift, the generator is formally the time-ordered exponential of the time-dependent generator.} This operator is well known in quantum mechanics and leads to what we call a \myemph{semi-group} structure for the evolution of $p(x,t)$ (or the wavefunction) as a result of the fact that $U(t) = U(s)U(s')$ for $t=s+s'$. For stochastic processes, this semi-group property \textit{is} the Markov property and $L^\dag$ is simply the generator of that semi-group.

The probability density of $X_t$ or the wavefunction of a quantum system are not the only objects whose evolution forms a semi-group. The expectation $\langle f(X_t)\rangle$ also does, since its evolution is linear, its generator being $L$.

In the 1940s, Mark Kac (pronounced \textit{khats}) showed that the general exponential functional of a Markov process $X_t$ defined as
\be
G(x,t) = \left\langle e^{\int_0^t c(X_s)ds}\right\rangle_x
\ee 
also has, amazingly, a semi-group structure, provided that the function $c$ is smooth enough and such that the expectation exists. Here the notation $\langle \cdot \rangle_x$ means that $X_t$ is started at $x$, that is, $X_0=x$ with probability 1. Specifically, he showed that
\be
\p_t G(x,t) = \cL_c G(x,t),
\ee
where $\cL_c = L+c$, $L$ being the generator of $X_t$. This linear PDE, coupled with the initial condition $G(x,0)=\langle e^0\rangle_x=1$, is what is called the \myemph{Feynman-Kac formula}.

There are many derivations of that formula, some of which based on analogies with quantum mechanics \cite{majumdar2005}. The simplest I could find is very similar to what we call in physics the Kramers-Moyal expansion (see \cite{reichl1980}) and proceeds by considering $G(x,t+dt)$ to write
\be
G(x,t+dt) = e^{c(x)dt} \left\langle e^{\int_{dt}^{t+dt} c(X_s)ds}\right\rangle_x .
\ee
The initial condition for the expectation in this formula does not match the start of the integral, so we should propagate $X_0$ to $X_{dt}$ with the SDE (\ref{eqsde1}) to obtain
\be
G(x,t+dt) = e^{c(x)dt}\int p(\xi)\, G(x+\xi,t)\, d\xi,
\ee
where $p(\xi)$ is the probability density of the Gaussian increment $\xi$ that takes us from $X_0=x$ to $X_{dt}=x+\xi$. By Taylor-expanding $G(x+\xi,t)$ to second order around $x$, and taking the expectation with respect to $\xi$, we finally arrive at the correct PDE.

To connect this result to our goal of calculating large deviations, we only need to notice that the generating function $\langle e^{TkA_T}\rangle$ entering in the definition of the SCGF $\lambda(k)$ in Eq.~(\ref{eqscgf1}) is a particular case of Kac's functional, at least for $g=0$, corresponding to $c(x) = kf(x)$ so that $\cL_c=\cL_k$ with a slight abuse of notations. Therefore, this generating function satisfies the Feynman-Kac equation, which enables us to write
\be
G(x,t) =\langle e^{tkA_t}\rangle_x = (e^{t\cL_k }1)(x),
\ee
where the propagator $e^{t\cL_k}$ acts on the initial condition $G(x,0)=1$ to yield some non-trivial function of $x$ at time $t$. From there, we can use the results of Appendix~\ref{appnh} about the spectral decomposition of non-Hermitian operators to expand the initial unit function in the eigenbasis of $\cL_k$. This yields, with the completeness relation (\ref{eqcomp1}) and the normalization (\ref{eqnorm2}) adopted,
\be
1= \sum_i r_k^{(i)}(x),
\ee
where $r_k^{(i)}$ denotes an eigenfunction of $\cL_k$ (not necessarily all real), so that
\be
G(x,t)=\sum_i e^{\zeta_i t} r_k^{(i)}(x),
\ee
where $\zeta_i$ denotes an eigenvalue of $\cL_k$. In the long-time limit, the eigenvalue with largest real part will dominate the sum (if there is a gap) and, thus, we arrive at the result mentioned in Sec.~\ref{secge} that the SCGF is the largest eigenvalue of $\cL_k$. 

The same result holds for $g\neq 0$, since there is also a Feynman-Kac formula for this case, one in fact that was never considered by Kac (see Exercise~\ref{ex8}).

\section*{Exercises}
\label{secex}

{\small
All exercises are rated, rather subjectivity, according to Knuth's logarithmic rating system whereby 00~=~immediate, 10 = simple, 20 = medium, 30 = moderately hard, 40 = term project, 50 = research problem. Feel free to contact me for questions or comments.

\begin{enumerate}[1.]
\item \exdiff{10}\label{ex1} The conservation of probability in quantum mechanics is expressed by the requirement that $H$ is Hermitian. What is the corresponding property of $L$ for Markov processes?

\item \exdiff{15}\label{ex2} Show that the stationary density of Kramers equation (\ref{eqkram1}), in the absence of external forces ($\phi_t=0$), is a Gibbs distribution involving the total energy $H(q,p) = p^2/(2m)+V(q)$. Calculate its associated stationary Fokker-Planck current. Does it vanish? Discuss the consequence of this result in view of the fact that equilibrium systems are supposed to have vanishing currents. Source: Chap.~10 of \cite{risken1996}.

\item \exdiff{15}\label{ex3} Prove that the stationary density of the gradient SDE (\ref{eqgradsde1}) is the Gibbs distribution (\ref{eqgibbs1}). Do we still have that stationary distribution when $\sigma$ is not proportional to the identity matrix? Calculate the associated stationary current.

\item \exdiff{20}\label{ex4} Show that the stationary density of the linear SDE (\ref{eqsdelin1}), with attractor at $x=0$, is a Gaussian distribution of the form
\be
p_s(x) = \sqrt{\frac{\det C}{2\pi}} \exp\left(-\frac{1}{2}x\cdot C x\right),
\ee
where $C$ is a symmetric, positive matrix. What is the equation satisfied by $C$ involving $M$ and $D$? What is the stationary current $J_s(x)$? When is this density Gibbsian?

\item \exdiff{20}\label{ex5} Prove that the generator $L$ of gradient SDEs is self-adjoint with respect to the following inner product
\be
\langle f,g\rangle_{p_s} = \int f(x)g(x)\, p_s(x)\, dx,
\ee
where $p_s$ is the Gibbs stationary distribution. What do you conclude for the spectrum of $L$?

\item \exdiff{20}\label{ex6} For gradient SDE (\ref{eqgradsde1}), $\langle Lf,f\rangle_{p_s}$ is a so-called Dirichlet form:
\be
\langle Lf,f\rangle_{p_s}=\langle f,Lf\rangle_{p_s} = -\frac{\eps^2}{2}\| \nabla f\|^2_{p_s},
\ee
where $\|\cdot\|_{p}$ is the norm weighted by $p$. Prove this result.

\item \exdiff{15}\label{ex6b} Show that the expectation of the empirical density $\rho_T(x)$ defined in (\ref{eqed1}) is the stationary density $p_s(x)$ when $X_t$ is ergodic. Explain why this must also be the most probable value of $\rho_T$ in the long-time limit. [Hint: What is the distribution of $\rho_T$? Does it satisfy an LDP?]

\item \exdiff{15}\label{ex7} Show that the expectation of the empirical current $J_T(x)$ defined in (\ref{eqec1}) is the stationary Fokker-Planck current.

\item \exdiff{20}\label{ex8} The form of the tilted generator $\cL_k$ is derived from the Feynman-Kac equation in Appendix~\ref{appfk} for the case $g=0$. Adapt the proof for the case $g\neq 0$ leading to the full expression of $\cL_k$ shown in~(\ref{eqlk1}).

\item \exdiff{10}\label{ex8b} Show that $l_{k=0}(x)=p_s(x)$ and $r_{k=0}(x)=1$ for all $x$.

\item \exdiff{15}\label{ex9} Show for gradient SDEs that $L p_s=p_sL^\dag$. Then use this result to show that $\cH_k$, as defined by the symmetrization (\ref{eqsym1}), is Hermitian.

\item \exdiff{10}\label{ex10} Derive for gradient SDEs the expression of the quantum generator $\cH_k$ shown in (\ref{eqh1}), together with the corresponding potential $V_k(x)$ shown in (\ref{eqv1}).

\item \exdiff{10}\label{ex10b} Derive the relations (\ref{eqt1}) and (\ref{eqt2}) between $\psi_k$, $r_k$, and $l_k$ using the relation (\ref{eqsym1}) between $\cH_k$ and $\cL_k$.

\item \exdiff{20}\label{ex13} Repeat the calculations of Sec.~\ref{secexou} for 
\be
V_T = \frac{1}{T}\int_0^T X_t^2 dt,
\ee
which represents the empirical variance of $X_t$. What is the related quantum problem? Find $\lambda(k)$, $r_k(x)$ and $l_k(x)$ (correctly normalized). Source: \cite{majumdar2002}.

\item \exdiff{45}\label{ex14} Repeat the previous exercise by replacing $X_t^2$ by $X_t^k$, $k>2$.


\item \exdiff{20}\label{ex14a} Study the large deviations of the entropy production $\Sigma_T$, as defined in Eq.~(\ref{eqep1}), of the Ornstein-Uhlenbeck process. Assume, first, that $X_0$ and $X_T$ have fixed values, e.g., $X_0=a$ and $X_T=b$. Then repeat the calculation for $X_0=a$, but by integrating $X_T$ over $\reals$. Finally, assume that $X_0$ is distributed according to $p_s(x)$. You should see in each case a very different effect of the ``boundary terms'' $X_0$ and $X_T$.

\item \exdiff{25}\label{ex14b} Combine the calculation of Sec.~\ref{secexou} with the previous exercise to find the rate function of
\be
Q_T = \frac{X_0^2}{T} -\frac{X_T^2}{T} +\frac{1}{T}\int_0^T X_t dt
\ee 
for the Ornstein-Uhlenbeck process. This observable can be related to the heat exchanged by a Brownian particle with its environment when manipulated by laser tweezers \cite{zon2004}. Obtain the rate function for the three cases considered before: i) $X_0=a$, $X_T=b$; ii) $X_0=a$ and $X_T\in\reals$; iii) $X_0\sim p_s(x)$ and $X_T\in \reals$. You should see that the region where $\lambda(k)<\infty$ is different in each case, although $\zeta_{\max}(\cL_k)$ is defined for all $k\in\reals$. [Hint: You will need the generating function of the $\chi^2$ distribution.]

\item \exdiff{30}\label{ex15} Calculate for the Ornstein-Uhlenbeck process the rate function $I(r)$ associated with
\be
r_T=\frac{1}{T}\int_0^T \chi_{[-1,1]}(X_t)\, dt.
\ee 
where $\chi_S(x)$ is the indicator function equal to 1 if $x\in S$ and $0$ otherwise. This observable represents the fraction of time $X_t$ spends in the interval $[-1,1]$, so it is a random variable taking values in $[0,1]$. What happens when $\gamma\ra 0$? Sources: \cite{angeletti2015,tsobgni2016b}.

\item \exdiff{40}\label{ex16} Show that the occupation fraction $r_T$ defined in the previous exercise can be obtained by integrating the empirical distribution $\rho_T(x)$ defined in (\ref{eqed1}) over $x\in [-1,1]$. Can you use this result to derive the rate function $I(r)$ from the known rate function(al) $I(\rho)$ of $\rho_T(x)$? Source: \cite{angeletti2015}.

\item \exdiff{25}\label{ex17} Repeat the exercise about the occupation time for $\chi_{[0,\infty)}(x)$ so as to study the fraction of time $X_t$ stays positive. Source: \cite{majumdar2005}.

\item \exdiff{20}\label{ex12} Consider the SDE
\be
d\theta_t = \gamma dt+\sigma dW_t,\qquad \theta_t \in [0,2\pi),
\ee
representing the motion of a Brownian particle on the unit circle with a drive or torque $\gamma$. Is this a gradient (equilibrium) system? What is the stationary Fokker-Planck current $J_s(x)$? Obtain the rate function $I(j)$ characterizing the fluctuations of the total empirical current
\be
J_T = \frac{1}{T}\int_0^T d\theta_t = \int J_T(x)dx.
\ee
What are the natural boundary conditions for $r_k(x)$ and $l_k(x)$ in this case? Can $\cL_k$ be symmetrized when $\gamma=0$? Source: \cite{tsobgni2016} and references therein.

\item \exdiff{25}\label{ex17b} Calculate the rate function of the entropy production $\Sigma_T$ for the 2D linear transversal SDE defined by
\be
F(x,y) = 
\left(
\begin{array}{cc}
-1 & -1 \\
1 & -1
\end{array}
\right)
\left(
\begin{array}{c}
x\\
y
\end{array}
\right)
\ee
and $\sigma = \eps\id$. Is this SDE gradient? Why is it called ``transversal''? What do you obtain if you try to symmetrize it? Source: \cite{noh2014}.

\item \exdiff{25} \label{ex17c} Show that the tilted generator associated with the entropy production $\Sigma_T$ satisfies the symmetry
\be
\cL_k^\dag = \cL_{-k-c} ,
\ee
where $c$ is some constant. Derive from this a symmetry satisfied by the SCGF and rate function. Such a symmetry is known as a \myemph{fluctuation relation}. Source: \cite{lebowitz1999}.

\item \exdiff{30}\label{ex18} Derive the expression of the tilted generator $\cL_k$ for SDEs in which the noise is multiplicative, that is, in which the noise matrix $\sigma$ depends on $x$. In this case, you must specify the stochastic convention used for interpreting the product $\sigma(x) dW_t$.

\item \exdiff{25} Another approach for deriving the SCGF is to discretize in time an SDE to obtain a Markov chain for which $A_T$ is then a sum. Use this approach to confirm that the SCGF $\lambda(k)$ is given by the dominant eigenvalue of $\cL_k$. Source: Chap.~V of \cite{bucklew1990} and the lecture notes of my large deviation course (see my website).

\item \exdiff{40}\label{ex19} Rewrite all the large deviations of these notes for Markov chains. Then do the same for Markov jump processes or continuous-time Markov chains. Source: \cite{chetrite2014}.

\item \exdiff{45}\label{ex20} Do the eigenvectors $r_k$ and $l_k$ have any probabilistic or physical interpretation? Sources: \cite{chetrite2013,chetrite2014}.

\item \exdiff{25} We have seen that, in the mapping to the quantum Hamiltonian, the noise parameter is essentially $\hbar$. With this, comment on the following: The low-noise limit of SDEs is equivalent to the semi-classical limit of quantum mechanics. Source: \cite{touchette2009}.
\end{enumerate}
}

\section*{In memoriam}

I dedicate this paper to the memory of E.~G.~D.\ Cohen (1923-2017): friend, collaborator, physicist and raconteur extraordinaire, who wisely told me once that the most important thing in science is to follow your heart.

\begin{acknowledgments}
I thank Francesco Coghi, Johan Du Buisson, Paul M. Geffert, Jan C.\ Louw, Thomas McGrath, Jan Meibohm, Takahiro Nemoto, Daniel Nickelsen, Audun Skaugen, and Shou-Wen Wang for comments on these notes. I also thank Marco Baiesi, Alberto Rosso, and Thomas Speck for the invitation to the 2017 School on Fundamental Problems in Statistical Physics. My trip there was funded by the National Research Foundation of South Africa (Grant no.\ 96199).
\end{acknowledgments}

\bibliography{masterbib}

\end{document}